# Language processing in the autistic brain: A meta-analysis of neuroimaging studies across modalities and age

Hanlin Wu, Bei Xiao, Qingxi He
Department of Linguistics and Modern Languages
The Chinese University of Hong Kong
Hong Kong

**Corresponding author:** Hanlin Wu (hanlinwu@cuhk.edu.hk)
**Running title:** Language processing in the autistic brain

## Abstract

Language is central to social communication, which is altered in autism, and language abilities vary widely among autistic individuals. How the autistic brain processes language is central to understanding autistic communication. Yet neuroimaging studies report inconsistent findings on how autistic brain activity is altered during language processing. We synthesized 45 fMRI and PET studies (753 autistic and 777 non-autistic participants) with seed-based *d* mapping with permutation of subject images (SDM-PSI). We estimated autistic and non-autistic activation and their difference with meta-regressions adjusted for each study's language modality and mean age. Autistic and non-autistic brains activated a similar left-lateralized language network. Autistic brains showed reduced activity throughout the left perisylvian cortex, from the inferior frontal gyrus to the mid superior temporal gyrus and the posterior superior temporal sulcus, and increased activity in the right temporal pole. Spoken language showed a larger autistic reduction than written language in the left mid superior temporal gyrus. Adults, but not children and adolescents, showed increased autistic activity in the right superior temporal gyrus and cuneus. Autistic language processing thus relies on the typical left-lateralized network but shows reduced leftward lateralization.

# 1. Introduction

Altered social communication is among the defining features of autism (American Psychiatric Association, 2013). Language is a central part of social communication, and language ability is among the strongest predictors of later social and educational outcomes in autism (Kjelgaard & Tager-Flusberg, 2001; Schaeffer et al., 2023). Autistic individuals vary widely in their language abilities: some only have minimal language (at most a few words or phrases), while others have language ability within the typical range (Boucher, 2012; Schaeffer et al., 2023). Even autistic individuals with typical language ability often use it atypically (Boucher, 2012; Schaeffer et al., 2023). How the autistic brain processes language is therefore central to understanding autistic communication. Functional neuroimaging has explored autistic language processing for nearly three decades. The literature is substantial but heterogeneous: some studies have described reduced left-hemisphere activity (e.g., Kana et al., 2006; Lai et al., 2012) and increased right-hemisphere activity (e.g., Kleinhans et al., 2008; Mason et al., 2008; Redcay & Courchesne, 2008), while others reported no group difference in whole-brain activity at all (e.g., Knaus et al., 2017; McCabe et al., 2025). Synthesis across studies is therefore needed to establish whether and how autistic brain activity during language processing differs from that of non-autistic brains.

## 1.1. Reduced left temporal activity in autism

Several syntheses have explored this question. Herringshaw et al. (2016) conducted an activation likelihood estimation (ALE) meta-analysis of language processing in autism across 22 fMRI and PET studies. They found autistic brain activity reduced in the left superior temporal gyrus (STG) and inferior frontal gyrus (IFG) and in the bilateral middle temporal gyri (MTG), but increased in the right STG and IFG and in the left lingual and middle frontal gyri. Philip et al. (2012) meta-analyzed fMRI studies across six task domains, and in the auditory and language domain autistic participants showed less activity than non-autistic participants in the bilateral STG and more activity in the right precentral gyrus and the left cerebellum. Both syntheses thus found reduced autistic brain activity in the lateral temporal cortex, but they reported increased activity in different regions.

More recent syntheses have focused on subdomains of language processing. Hua et al. (2024) synthesized nine studies of auditory language comprehension in pediatric autistic samples and reported reduced autistic brain activity in the right STG and in the left MTG extending into the left insula, with no region of increased brain activity. Cheng et al. (2026) synthesized six studies of figurative language and reported a single cluster of reduced autistic brain activity spanning the left STG and MTG. Duvall et al. (2023) synthesized studies of theory of mind and pragmatic communication, and in the 13 studies of pragmatic communication they found reduced autistic brain activity mainly in the left hemisphere (the left inferior frontal, superior and middle temporal and medial prefrontal cortex). Phan et al. (2021) synthesized 11 studies of semantic processing and found that the autistic brain activity lay entirely in the left hemisphere and the non-autistic activity in both hemispheres. Tryfon et al. (2018) synthesized 18 studies of speech perception and found no significant group difference in either direction. Most of the previous syntheses found reduced autistic brain activity in the left temporal cortex, but they reported inconsistent findings on whether the reduction extended to the left inferior frontal cortex and on whether autistic brain activity was increased in any brain regions.

### 1.2. Language processing differs between modalities and changes with age

Where and how autistic brain activity is altered may depend on the modality of the language and on the age of the participants. Spoken and written language enter the language network through different sensory cortices. Speech activates the auditory cortex of the bilateral STG while reading activates the ventral occipitotemporal cortex, and beyond these sensory cortices the two share a largely left-lateralized temporal, frontal and parietal network (Price, 2012). Autistic perception has been proposed to be enhanced for stimuli that primary sensory cortex can process on its own, such as pure tones, and reduced for stimuli that require later stages of the sensory hierarchy (the enhanced perceptual functioning account; Mottron et al., 2006). Speech is a sound with complex temporal and spectral structure (Samson et al., 2011); if the account holds, the group difference should be larger for heard than for read language and should already appear in the auditory cortex.

Age may matter as well. The non-autistic language network changes between childhood and adulthood. In a meta-analysis of spoken language comprehension, convergence in children was more bilateral in the temporal cortex than in adults and centered in the inferior frontal cortex on the pars triangularis, compared with the pars opercularis in adults (Enge et al., 2020). Leftward lateralization is established by about five years of age and increases only slightly thereafter, in frontal regions and in close relation to task performance (Weiss-Croft & Baldeweg, 2015). How age modulates the group difference is not established. Some differences between autistic and non-autistic brains have been described as largest in childhood: brain size in the first years of life (Redcay & Courchesne, 2005) and task-related brain activity (Dickstein et al., 2013). Philip et al. (2012) analyzed the pediatric and the adult studies of their auditory and language domain separately and found reduced autistic superior temporal activity in the pediatric studies only. However, a magnetoencephalography study of six language-impaired autistic and six non-autistic children listening to vowels suggests the opposite pattern: older autistic children showed more rightward lateralization and older non-autistic children more leftward lateralization, so that the group difference may increase with age (Flagg et al., 2005).

Both modality and age vary within and between the previous syntheses. Herringshaw et al. (2016) combined 8 studies of heard and 13 of read language, with autistic participants aged 7 to 52 years. Tryfon et al. (2018) and Hua et al. (2024) included only studies of spoken language, but the samples of Tryfon et al. (2018) ranged in mean age from 32 months to 34 years. If the group difference depends on modality or age, the findings of a synthesis would vary with its mix of studies, and syntheses with different mixes would produce inconsistent results. Restricting a synthesis to one subdomain, as the more recent syntheses did, removes part of this heterogeneity but divides an already small literature into smaller sets. The five recent syntheses based their group comparisons on 3 to 13 studies (Cheng et al., 2026; Duvall et al., 2023; Hua et al., 2024; Phan et al., 2021; Tryfon et al., 2018), fewer than the 17 to 20 experiments recommended for ALE (V. I. Müller et al., 2018).

### 1.3. The present study

We synthesized 45 fMRI and PET studies of language processing in autistic and non-autistic participants with seed-based *d* mapping with permutation of subject images (SDM-PSI; Albajes-Eizagirre et al., 2019). In each of three meta-analyses, one meta-regression estimated the effect with the modality of the language (spoken or written) and the mean age of the sample as moderators. We first described each group's language network, how it depends on modality and how it changes with age. We then tested where the groups differ and whether the group difference depends on modality or age.

Reduced leftward lateralization of language is common in autism (Lindell & Hudry, 2013). In autistic adults reading sentences, it was suggested to reflect increased right-hemisphere activity (Jouravlev et al., 2020). If reduced lateralization takes this form across studies, we expect autistic brain activity to be increased in the right-hemisphere counterparts of the left-hemisphere language regions, as Herringshaw et al. (2016) found in the right IFG and STG. Within the left hemisphere, the underconnectivity theory of autism proposes that autistic brains rely less on frontal and more on posterior processing (Just et al., 2004, 2012). Studies in several cognitive domains, including language, have found reduced frontal and increased posterior activity in autistic participants (see Just et al., 2012, for a review). In autistic participants reading sentences, Just et al. (2004) found reduced left inferior frontal and increased left posterior temporal activity. If the theory holds across studies, we expect autistic brain activity to be reduced in the left inferior frontal cortex and increased in the left posterior temporal cortex.

SDM-PSI recreates each study's map of effect sizes from its peak coordinates and their *t* values, imputing the unreported values within the bounds set by the study's statistical threshold. The studies' maps are combined by random-effects meta-regression, in which several moderators can be estimated together and their effects tested across the whole brain.

A comparison of the two groups' activation maps need not produce the same result as a synthesis of the group contrasts that the studies themselves report, because a region can be active in both groups and still be less active in one of them. Apparent differences between two groups' maps were often not significant in the direct contrast between the groups within single studies (Philip et al., 2012). We therefore treat the meta-analysis of the between-group contrasts as the test of a group difference, and the meta-analyses of each group's activation as descriptions of the language network of each group (Box 1).

**Box 1.** Terms used in this paper.

| Term | Meaning |
|---|---|
| Hedges' *g* | The standardized effect at a voxel, with Hedges' small-sample correction: in a within-group meta-analysis, the mean difference between language and control tasks divided by its standard deviation across participants; in the between-group meta-analysis, the difference between the autistic and the non-autistic group in that contrast divided by their pooled standard deviation (positive: autistic > non-autistic). |
| Within-group meta-analysis | The meta-analysis of the language > control contrasts of one group, autistic or non-autistic. It describes each group's language network. |
| Between-group meta-analysis | The meta-analysis of the group contrasts the studies report, in both directions, including studies that found no significant difference. It tests the group difference. |
| Moderator | A study characteristic in the meta-regression: modality (written or spoken) and mean age in years. The two moderators are centered on their means across the studies and estimated together, so that the effect of each moderator is adjusted for the other moderator. |
| Intercept map | The effect predicted with age and modality at their means across the studies (the mean age and the mix of spoken and written studies): the main result of each meta-analysis. |
| Moderator map | The change in the effect with one moderator: from written to spoken language, or per year of age. In the between-group meta-analysis, a moderator effect is a change in the group difference: a group-by-modality or group-by-age interaction. |
| Marginal mean | The effect predicted at one level of a moderator: spoken or written language with age at its mean, or the mean age of the pediatric or adult samples with modality at its mean. It describes a moderator effect. |
| Null study | A study whose whole-brain group test found no significant voxel in either direction. |

## 2. Methods

### 2.1. Search, selection and extraction

The review is reported according to the PRISMA 2020 guideline (Page et al., 2021) and was designed with reference to the recommendations of V. I. Müller et al. (2018) for neuroimaging meta-analysis. We searched Web of Science, PubMed and Scopus on 4 January 2026 for records published during the past 30 years (1 January 1996 to 31 December 2025). The search crossed four term blocks (autism, language, processing, and neuroimaging), with review articles excluded by document type; the full search strings are listed in Supplementary Material S1. The search returned 3,727 records (Web of Science 1,091, PubMed 1,352, Scopus 1,284). Duplicates (1,344), records without an abstract (49) and non-English records (32) were removed, and 2,302 records remained for title and abstract screening, after which 55 reports were assessed in full text. We also examined the studies included in or cited by seven previous syntheses of language in autism (Cheng et al., 2026; Herringshaw et al., 2016; Hua et al., 2024; Larson et al., 2025; Li et al., 2022; Phan et al., 2021; Tryfon et al., 2018). This check found eight reports that the database search had missed. With these, 63 reports were assessed in full text (Figure 1; Supplementary Tables S1 and S1b).

Reports were eligible if they met seven criteria: (1) an fMRI or PET study in which a language task was contrasted with a control condition; (2) an autistic group and a non-autistic comparison group, with each group's activation reported separately or the contrast between them; (3) at least one eligible contrast analyzed over the whole brain; (4) peak coordinates in a standard space (MNI or Talairach); (5) participants scanned awake; (6) the mean age of each group reported; and (7) results that could enter at least one of the three meta-analyses. Results restricted to anatomical regions of interest, to small-volume corrections or network masks, or to a functional mask derived from the same data (for example, a group test within the voxels active in either group) were not eligible and were excluded. Each excluded report received one primary reason (Figure 1; Supplementary Table S1). Of the 63 reports assessed in full text, 45 were included.

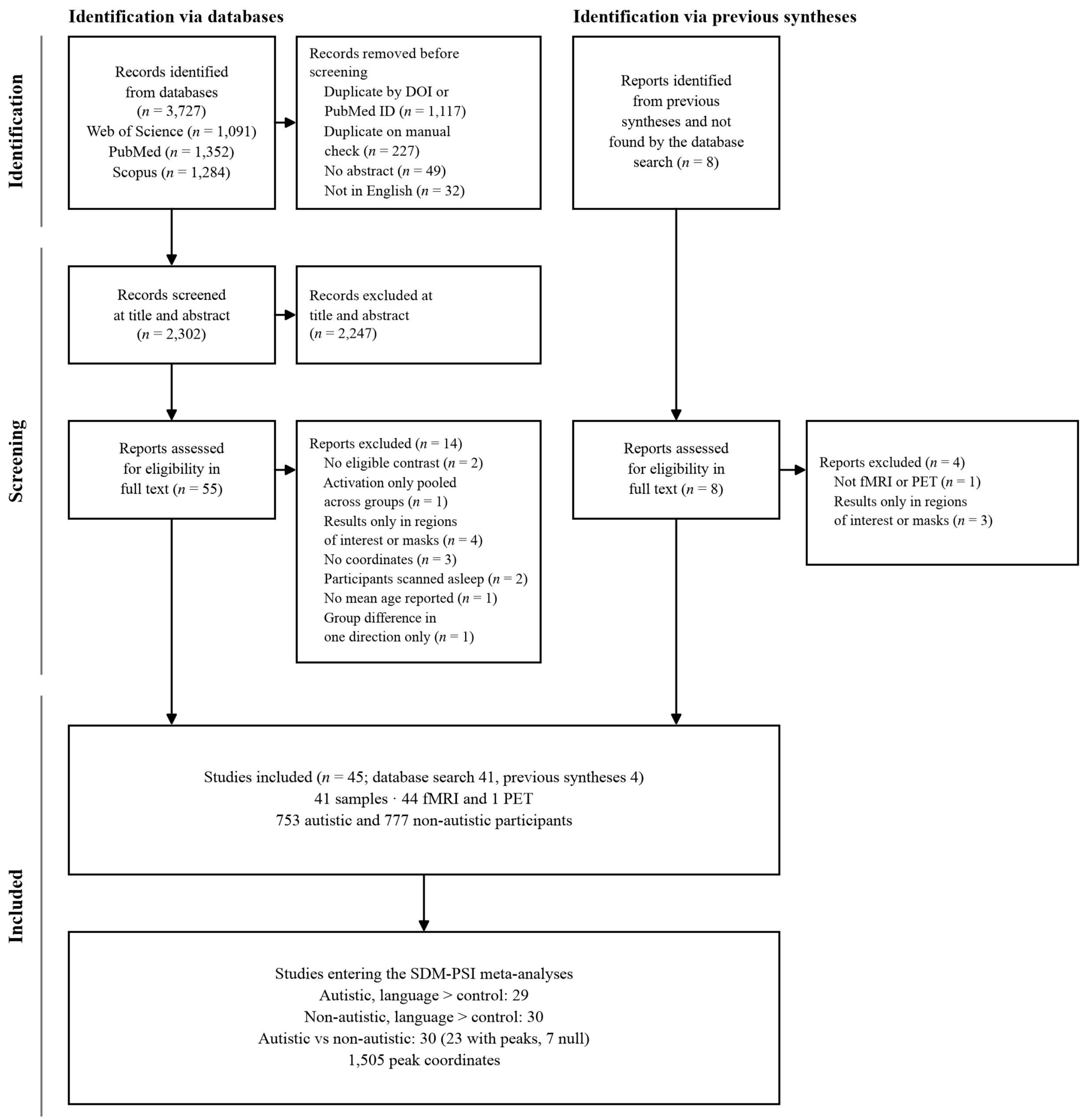

**Figure 1.** PRISMA 2020 flow diagram of the search, screening and selection of studies.

From the 45 included studies we extracted the peak coordinates, test statistics and sample sizes. We then checked five properties: that every coordinate fell inside the brain after conversion; that hemisphere labels agreed with the sign of *x*; that every statistic reached the study's threshold; that sample sizes were constant across a sample's contrasts; that no peak was entered twice within a contrast. Values were kept as the source printed them, except in one study (R.-A. Müller et al., 1999) where the peak table printed every lateral *x* coordinate with the sign opposite to the hemisphere label and the text followed the labels; we reversed these signs. Across the 45 included studies, 1,807 peaks were extracted, and 1,505 entered the three meta-analyses.

## 2.2. Coding

For each sample we coded language > control contrasts within the autistic group and within the non-autistic group, and group differences in such contrasts in each direction (autistic > non-autistic and non-autistic > autistic), with null results. Modality is the channel of the language the participants processed. Written studies included reading words, sentences and texts, reading aloud and naming from written descriptions. Spoken studies included perceiving speech (heard, audiovisual or lip-read), repetition and verbal fluency. Age is the mean age of each group in years or the *n*-weighted mean where only subgroups were reported. The autistic group's mean age entered the autistic meta-analysis; the non-autistic group's mean age entered the non-autistic meta-analysis; the *n*-weighted mean of both groups entered the between-group meta-analysis.

## 2.3. Samples and studies

Each participant sample entered each meta-analysis once. Where a sample contributed several eligible contrasts to one meta-analysis, we pooled their peaks, with identical coordinates entered once. For studies that stated that their participants overlapped, in full or in part, we merged them into one sample: Tesink et al. (2009, 2011), Gaffrey et al. (2007) and Shen et al. (2012), Chen et al. (2016) and Fan et al. (2021), and Graves et al. (2022) and McCabe et al. (2025). Where both studies of a merged sample entered the same meta-analysis, the sample took the *n* and mean ages of the study that analyzed more participants. The 45 studies thus describe 41 samples, and each sample counts as one study in the meta-analyses.

Every sample with a whole-brain group test entered the between-group meta-analysis when both directions of the test were settled by peaks, or by the study's statement that no group difference was found. A sample whose group test found no significant voxel in either direction entered as a null study.

Coordinates reported in Talairach space were converted to MNI space with the Lancaster transforms (Lancaster et al., 2007). For SDM-PSI, *t* values were kept as printed and *Z* values were converted to *t* with the degrees of freedom of the test. Peaks reported without a usable statistic entered with their sign only, and SDM-PSI estimated their effect sizes. Each study's threshold was taken from its voxel-level or cluster-forming threshold (Supplementary Material S1.3).

### 2.4. SDM-PSI meta-analyses

We ran three meta-analyses in SDM-PSI version 6.23 (Albajes-Eizagirre et al., 2019): autistic language > control, non-autistic language > control and the signed between-group difference. Each study's map of effect sizes (Hedges' *g*) was recreated on SDM's gray-matter template at 2 mm with a fully anisotropic kernel (SDM's default full width at half maximum of 20 mm; Radua et al., 2014), and the unknown values were imputed 50 times within the bounds set by the study's threshold. In every meta-analysis the imputed maps were combined by random-effects meta-regression with modality and age as moderators, and the results of the 50 imputations combined by Rubin's rules (Rubin, 1987). SDM-PSI centers each moderator on its mean across the studies of the analysis, so the intercept is the effect predicted with age and modality at their means. The intercept and each moderator were tested by 1,000 permutations of subject images, with threshold-free cluster enhancement (TFCE; Smith & Nichols, 2009). Maps were thresholded at a TFCE family-wise-error (FWE) corrected $p < 0.05$ in each direction, with a minimum extent of 10 voxels.

For each cluster we report the peak (the voxel with the largest |SDM-*Z*|), its effect (*g* for the intercept, the coefficient for a moderator) with its 95% confidence interval, its TFCE-FWE *p* and its heterogeneity ($I^2$; Higgins & Thompson, 2002). Local maxima within a cluster were at least 8 mm apart. For every cluster of the between-group intercept map we also counted the studies that report a peak inside it or within 10 mm of its edge (Figure 3B): these studies inform the cluster most directly. At each intercept peak we tested for small-study effects with an Egger-type test (Egger et al., 1997) and for an excess of significant findings (Ioannidis & Trikalinos, 2007). Funnel plots are provided in Supplementary Figure S4.

### 2.5. Marginal means

To describe each significant moderator effect, we computed at every voxel the effect predicted at each level of the moderator (level maps), in the manner of estimated marginal means. The modality levels were spoken and written language, with age at its mean; the age levels were the mean age of the pediatric samples (under 18 years) and that of the adult samples (18 years or over), with modality at its mean. SDM-PSI's permutation test applies to the intercept or a moderator coefficient but not to the effect predicted at a given level of a moderator, which combines the two. These level maps are therefore thresholded at an uncorrected voxel-level $p < 0.001$ with at least 10 voxels and are descriptive. We report the predicted effect with its 95% confidence interval at every moderator peak and show the between-group level maps as a description in the supplement (Supplementary Figure S5).

### 2.6. Anatomical labeling

Peaks were labeled with the Automated Anatomical Labelling atlas 3 (AAL3; Rolls et al., 2020), as the region holding the largest share of a 5 mm sphere around the peak. Because a gyral atlas such as AAL3 has no parcel for a sulcus, the choice between the superior temporal sulcus and a gyral label was made with the HCP-MMP1 parcellation (Glasser et al., 2016), which divides the superior temporal sulcus along its length and between its banks: a peak was named by the sulcus if at least half of the labeled voxels of the sphere fell in the four sulcal parcels of HCP-MMP1. Superior temporal labels take an anterior, mid or posterior qualifier from the $y$ coordinate of the peak (posterior $y \leq -30$, anterior $y \geq 0$, mid otherwise).

## 3. Results

### 3.1. Characteristics of included studies

The 45 included studies (44 fMRI and one PET), published between 1999 and 2025, describe 41 samples with 753 autistic and 777 non-autistic participants (Table 1). The autistic meta-analysis comprised 29 studies with 515 participants and 607 peaks, the non-autistic meta-analysis comprised 30 studies with 537 participants and 686 peaks, and the between-group meta-analysis comprised 30 studies with 546 autistic and 561 non-autistic participants, 23 with peaks (212 peaks) and 7 null studies. Spoken studies made up 18 of the 29 autistic, 19 of the 30 non-autistic and 16 of the 30 between-group studies. The mean age of the samples was 20.5 ± 7.4 years in the autistic meta-analysis (study means 10.9 to 36.0), 20.4 ± 7.8 in the non-autistic meta-analysis (study means 10.6 to 34.0) and 20.8 ± 8.0 in the between-group meta-analysis (study means 10.1 to 36.6). Studies of pediatric samples (under 18 years) made up 13 of the 29 autistic, 14 of the 30 non-autistic and 13 of the 30 between-group studies. Modality and age were nearly uncorrelated ($|r| < 0.16$ in each meta-analysis).

Heterogeneity was low at the intercept peaks ($I^2$ from 0.1% to 25.1%) and ranged from 0.3% to 54.8% at the moderator peaks (Tables 2 and 3). The Egger-type test found no small-study effect at any intercept peak (all $p \geq 0.74$), and the test for excess significance found no excess of significant studies (all $p \geq 0.69$; Supplementary Figure S4).

**Table 1.** Characteristics of included studies.

| Study | *n* | Mean age | Modality | Autistic | Non-autistic | Between-group |
|---|---|---|---|---|---|---|
| R.-A. Müller et al., 1999 | 5 / 5 | 26.6 / 27.6 | Spoken | ✓ | ✓ | |
| Harris et al., 2006 | 14 / 22 | 36.0 / 31.0 | Written | ✓ | ✓ | |
| Kana et al., 2006 | 12 / 13 | 22.5 / 20.3 | Written | | | ✓ |
| Wang et al., 2006 | 18 / 18 | 11.9 / 11.9 | Spoken | ✓ | ✓ | |
| Gaffrey et al., 2007‡ | 10 / 10 | 26.1 / 25.3 | Written | ✓ | ✓ | ✓ |
| Wang et al., 2007 | 18 / 18 | 12.5 / 11.8 | Spoken | ✓ | ✓ | |
| Kleinhans et al., 2008 | 14 / 14 | 23.8 / 22.4 | Spoken | ✓ | ✓ | ✓ |
| Knaus et al., 2008 | 12 / 12 | 15.5 / 14.9 | Written | ✓ | ✓ | |
| Mason et al., 2008 | 18 / 18 | 26.5 / 27.4 | Written | ✓ | ✓ | |
| Tesink et al., 2009† | 24 / 24 | 26.3 / 26.2 | Spoken | ✓ | ✓ | Null |
| Anderson et al., 2010 | 26 / 15 | 21.7 / 22.5 | Spoken | ✓ | ✓ | |
| Groen et al., 2010 | 16 / 26 | 15.3 / 15.7 | Spoken | ✓ | ✓ | ✓ |
| Hesling et al., 2010 | 8 / 8 | 23.4 / 23.1 | Spoken | ✓ | ✓ | ✓ |
| Scott-Van Zeeland et al., 2010 | 18 / 18 | 12.6 / 11.6 | Spoken | ✓ | ✓ | ✓ |
| Catarino et al., 2011 | 12 / 12 | 27.0 / 34.0 | Written | ✓ | ✓ | Null |
| Mizuno et al., 2011 | 15 / 15 | 24.7 / 24.7 | Written | ✓ | ✓ | ✓ |
| Tesink et al., 2011† | 24 / 24 | 26.3 / 26.2 | Spoken | ✓ | ✓ | Null |
| Eigsti et al., 2012 | 16 / 11 | 13.7 / 13.7 | Spoken | ✓ | ✓ | ✓ |
| Hubbard et al., 2012 | 10 / 10 | 13.1 / 12.1 | Spoken | ✓ | ✓ | |
| Kana & Wadsworth, 2012 | 16 / 16 | 20.0 / 21.6 | Written | | | ✓ |

| **Study** | ***n*** | **Mean age** | **Modality** | **Autistic** | **Non-autistic** | **Between-group** |
|---|---|---|---|---|---|---|
| Lai et al., 2012 | 12 / 21 | 12.4 / 10.7 | Spoken | | ✓ | ✓ |
| Shen et al., 2012‡ | 14 / 14 | 24.1 / 24.2 | Written | ✓ | ✓ | ✓ |
| Doyle-Thomas et al., 2013 | 18 / 16 | 14.9 / 14.7 | Spoken | ✓ | ✓ | ✓ |
| Gebauer et al., 2014 | 19 / 20 | 26.2 / 24.5 | Spoken | | | ✓ |
| Han et al., 2014 | 10 / 10 | 16.0 / 14.9 | Written | | | ✓ |
| Moseley et al., 2015 | 18 / 18 | 30.4 / 28.6 | Written | | | ✓ |
| Chen et al., 2016§ | 37 / 35 | 13.3 / 13.3 | Written | ✓ | ✓ | ✓ |
| Murdaugh et al., 2016 | 26 / 19 | 10.9 / 10.6 | Written | ✓ | ✓ | ✓ |
| Schelinski et al., 2016 | 16 / 16 | 33.8 / 33.7 | Spoken | ✓ | ✓ | ✓ |
| Chouinard et al., 2017 | 12 / 12 | 32.4 / 33.0 | Spoken | ✓ | ✓ | ✓ |
| Kana et al., 2017 | 15 / 15 | 21.6 / 27.3 | Written | ✓ | ✓ | ✓ |
| Knaus et al., 2017 | 15 / 19 | 13.6 / 12.6 | Written | ✓ | ✓ | Null |
| Green et al., 2018 | 15 / 16 | 14.1 / 15.0 | Spoken | ✓ | ✓ | |
| Kim et al., 2018 | 15 / 18 | 9.7 / 10.5 | Written | | | ✓ |
| Tietze et al., 2019 | 16 / 16 | 39.5 / 33.8 | Spoken | | | ✓ |
| Hernandez et al., 2020 | 26 / 28 | 13.8 / 13.8 | Spoken | ✓ | ✓ | |
| Fan et al., 2021§ | 61 / 67 | 12.7 / 12.8 | Written | ✓ | ✓ | ✓ |
| I.-F. Lin et al., 2021 | 21 / 24 | 27.9 / 27.8 | Spoken | | | Null |
| Graves et al., 2022¶ | 19 / 22 | 20.8 / 21.3 | Written | ✓ | ✓ | ✓ |
| Leipold et al., 2023 | 22 / 21 | 10.7 / 10.7 | Spoken | | | ✓ |
| O’Brien et al., 2023 | 23 / 28 | 12.2 / 11.9 | Spoken | | | Null |
| Schelinski & von Kriegstein, 2023 | 17 / 17 | 30.5 / 31.4 | Spoken | ✓ | ✓ | Null |
| Ross et al., 2024 | 41 / 41 | 19.5 / 18.9 | Spoken | ✓ | ✓ | |
| H.-T. Lin et al., 2025 | 54 / 56 | 12.3 / 11.9 | Written | | | Null |
| McCabe et al., 2025¶ | 19 / 21 | 20.3 / 21.7 | Written | ✓ | ✓ | ✓ |

*Note.* *n* and mean age (years) are shown as autistic / non-autistic. Studies whose participants overlap, as the studies state, enter each meta-analysis as one sample: † Tesink et al. (2009, 2011), one experiment; ‡ Gaffrey et al. (2007) and Shen et al. (2012), the latter including the ten autistic participants of the former; § Chen et al. (2016) and Fan et al. (2021), which share fewer than 40% of each group; ¶ Graves et al. (2022) and McCabe et al. (2025), from the same scanning sessions. Fan et al. (2021) report two age subgroups; their *n* are summed and their mean ages weighted by *n*. For Lai et al. (2012), 21 controls entered the non-autistic meta-analysis and the group test compared 12 age-matched controls with the 12 autistic participants scanned while alert; the autistic map included sedated participants and was not entered.

### 3.2. Language networks in autistic and non-autistic brains

Language activated a left-lateralized network in both groups (Figure 2A; Table 2). In the autistic group, almost all of the activation formed one cluster (563,728 $mm^3$). It peaked in the left IFG pars triangularis ($g$ = 0.60, 95% CI 0.48 to 0.72) and spanned the perisylvian cortex of both hemispheres and the medial frontal cortex, with local maxima in the left MTG, the left posterior superior temporal sulcus (STS) and the left supplementary motor area (SMA). A second, small cluster lay in the right angular gyrus (160 $mm^3$; $g$ = 0.34). In the non-autistic group, the larger cluster (559,688 $mm^3$) also peaked in the left IFG pars triangularis ($g$ = 0.64, 95% CI 0.53 to 0.76) and spanned the perisylvian cortex of both hemispheres. The second (92,184 $mm^3$) peaked in the left medial superior frontal gyrus ($g$ = 0.55) and extended over the medial frontal cortex.

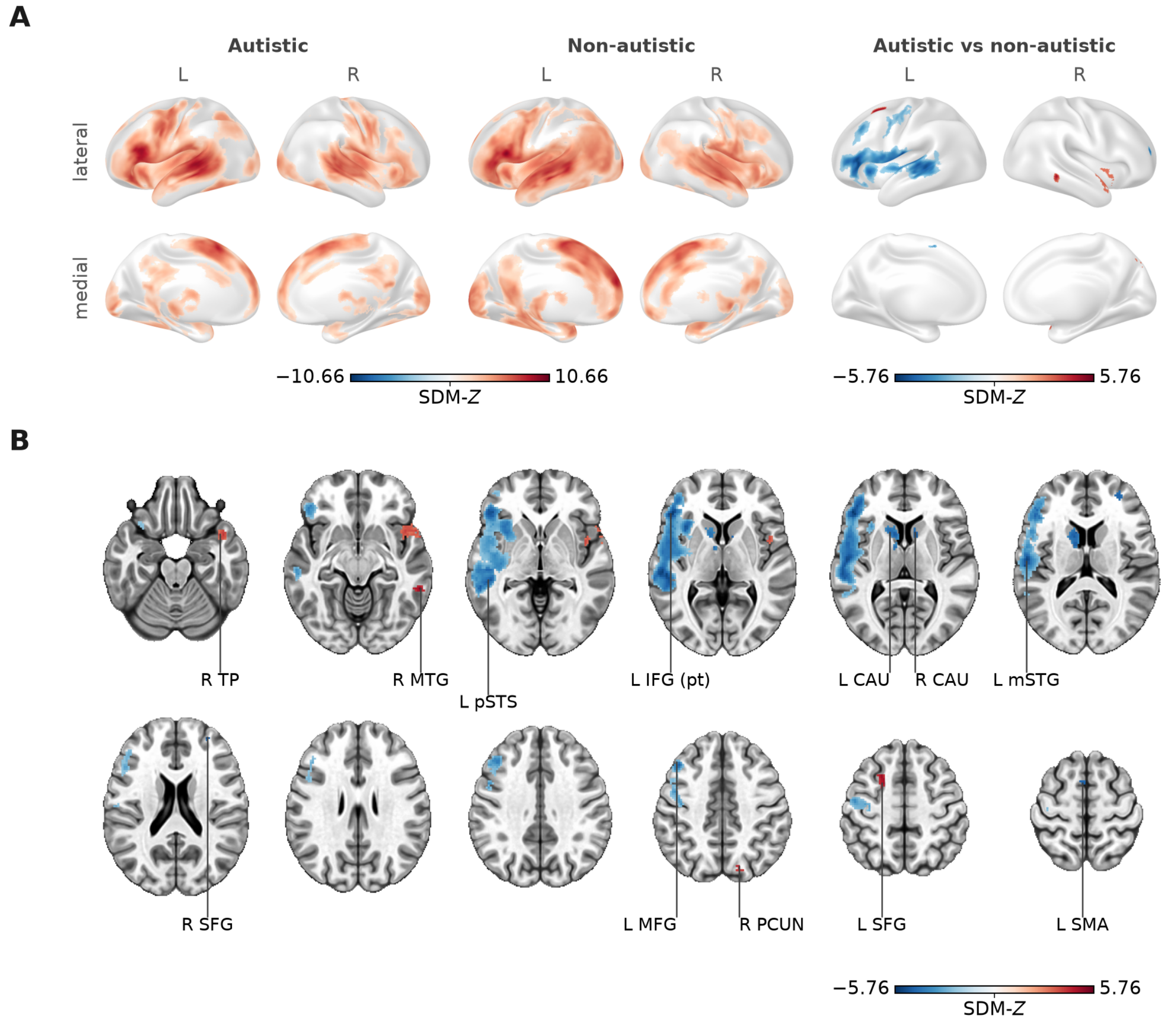

**Figure 2.** Language networks in autistic and non-autistic groups and the difference between the groups. (A) In the within-group maps, red marks activation for language; in the between-group map, red marks autistic > non-autistic and blue non-autistic > autistic. (B) The between-group map on axial slices ($z$ = −22, −10, −2, 4, 8, 14, 20, 28, 36, 44, 52 and 60 mm). L = left; R = right; TP = temporal pole; MTG = middle temporal gyrus; pSTS = posterior superior temporal sulcus; IFG (pt) = inferior frontal gyrus pars triangularis; CAU = caudate nucleus; mSTG = mid superior temporal gyrus; SFG = superior frontal gyrus; MFG = middle frontal gyrus; PCUN = precuneus; SMA = supplementary motor area. TFCE family-wise-error $p < 0.05$ (1,000 permutations) with at least 10 voxels. Labeled slices of the within-group maps are in Supplementary Figure S1.

**Table 2.** Clusters of the within-group meta-analyses.

| **Map** | **Direction** | **Peak region** | **Peak MNI** | **Voxels (mm³)** | **SDM-*Z*** | ***g* or coefficient [95% CI]** | $p_{FWE}$ | **I² (%)** |
|---|---|---|---|---|---|---|---|---|
| **Autistic** | | | | | | | | |
| Intercept | activation | L IFG (pt) | −50, 22, 22 | 70,466 (563,728) | 9.97 | 0.60 [0.48, 0.72] | 0.001 | 3.6 |
| Intercept | activation | R ANG | 34, −58, 50 | 20 (160) | 5.48 | 0.34 [0.22, 0.46] | 0.036 | 1.3 |
| Modality | higher for spoken language | R RolOp | 46, −12, 14 | 6,398 (51,184) | 5.43 | 0.71 [0.45, 0.96] | 0.001 | 16.7 |
| Modality | higher for spoken language | L pSTG | −42, −32, 12 | 2,824 (22,592) | 4.34 | 0.60 [0.33, 0.87] | 0.001 | 23.9 |
| Modality | higher for spoken language | L TP | −38, 14, −32 | 19 (152) | 2.65 | 0.34 [0.09, 0.59] | 0.050 | 7.9 |
| Age | increases with age | L INS | −30, 24, 2 | 418 (3,344) | 4.81 | 0.044 [0.026, 0.062] | 0.001 | 11.4 |
| Age | increases with age | L CB VI | −30, −52, −30 | 389 (3,112) | 3.58 | 0.035 [0.016, 0.054] | 0.004 | 27.2 |
| Age | decreases with age | R aSTG | 62, 0, −14 | 448 (3,584) | −2.61 | −0.024 [−0.043, −0.006] | 0.007 | 9.6 |
| Age | decreases with age | L TP | −38, 16, −32 | 16 (128) | −2.59 | −0.024 [−0.042, −0.006] | 0.033 | 6.5 |
| **Non-autistic** | | | | | | | | |
| Intercept | activation | L IFG (pt) | −54, 18, 18 | 69,961 (559,688) | 10.66 | 0.64 [0.53, 0.76] | 0.001 | 5.3 |
| Intercept | activation | L mSFG | −6, 58, 22 | 11,523 (92,184) | 9.12 | 0.55 [0.43, 0.67] | 0.001 | 6.7 |

| **Map** | **Direction** | **Peak region** | **Peak MNI** | **Voxels (mm³)** | **SDM-*Z*** | ***g* or coefficient [95% CI]** | $p_{FWE}$ | **I² (%)** |
|---|---|---|---|---|---|---|---|---|
| Modality | higher for spoken language | R mSTS | 60, −12, −12 | 3,975 (31,800) | 4.83 | 0.59 [0.35, 0.83] | 0.001 | 4.7 |
| Modality | higher for spoken language | L MTG | −64, −18, 0 | 3,930 (31,440) | 4.50 | 0.83 [0.47, 1.19] | 0.001 | 54.8 |
| Modality | higher for written language | R SMA | 6, 24, 52 | 283 (2,264) | −3.67 | −0.57 [−0.87, −0.26] | 0.012 | 30.7 |
| Age | increases with age | R CB VIII | 12, −68, −48 | 3,689 (29,512) | 3.50 | 0.034 [0.015, 0.053] | 0.001 | 30.8 |

*Note.* TFCE family-wise-error $p < 0.05$ (1,000 permutations), at least 10 voxels on a 2-mm grid. Values are at the cluster peak (largest |SDM-*Z*|). For the intercept, the effect is Hedges' *g*; for the moderators, it is the coefficient: the difference in *g* between spoken and written language (modality) or the change in *g* per year of mean age (age). $I^2$ is at the peak. IFG = inferior frontal gyrus; pt = pars triangularis; a/m/pSTG and a/m/pSTS = anterior/mid/posterior superior temporal gyrus and sulcus; MTG = middle temporal gyrus; TP = temporal pole; INS = insula; RolOp = Rolandic operculum; mSFG = medial superior frontal gyrus; SMA = supplementary motor area; ANG = angular gyrus; CB = cerebellum.

### 3.3. Group differences

Autistic brain activity was reduced mainly in the left hemisphere and increased mainly in the right (Figures 2 and 3; Table 3). The largest cluster of reduced autistic activity (43,920 $mm^3$) peaked in the left mid STG ($g = -0.47$, 95% CI −0.63 to −0.31) and extended from the left IFG pars triangularis through the frontal operculum and the anterior insula to the mid and posterior STG and the posterior STS, and up to the middle frontal gyrus and the precentral gyrus. Its largest local maxima outside the mid STG were in the left IFG pars triangularis ($g = -0.47$), the anterior insula ($g = -0.42$), the posterior STS ($g = -0.43$) and the middle frontal gyrus ($g = -0.44$). Autistic activity was also reduced in the bilateral caudate nuclei (left 1,856 $mm^3$, $g = -0.39$; right 200 $mm^3$, $g = -0.40$) and in two small frontal clusters: the right superior frontal gyrus (240 $mm^3$; $g = -0.38$) and the left SMA (104 $mm^3$; $g = -0.39$). It was increased in the right temporal pole and the adjacent anterior insula (3,376 $mm^3$; $g = 0.35$, 95% CI 0.19 to 0.52) and in three small clusters: the left superior frontal gyrus (536 $mm^3$; $g = 0.44$), the right posterior MTG (368 $mm^3$; $g = 0.44$) and the right precuneus (168 $mm^3$; $g = 0.35$).

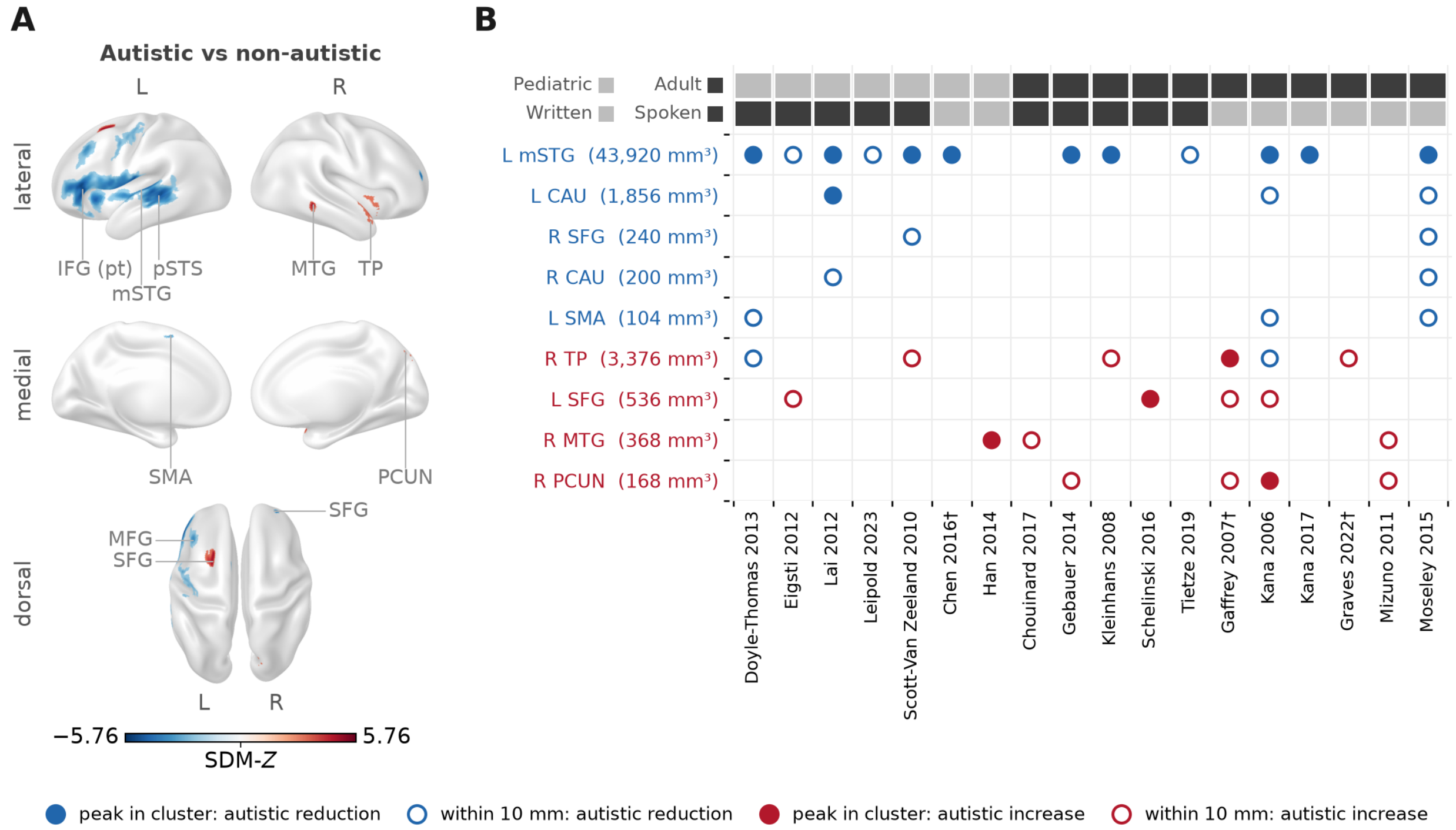

**Figure 3.** The group difference and the studies near each cluster. (A) The between-group intercept map in lateral, medial and dorsal views. (B) The studies that reported a peak inside each between-group cluster (filled circle) or within 10 mm of its edge (open circle). Blue: reduced autistic activity; red: increased autistic activity. † Two studies of one sample. IFG (pt) = inferior frontal gyrus pars triangularis; mSTG = mid superior temporal gyrus; pSTS = posterior superior temporal sulcus; MTG = middle temporal gyrus; TP = temporal pole; SMA = supplementary motor area; PCUN = precuneus; MFG = middle frontal gyrus; SFG = superior frontal gyrus; CAU = caudate nucleus; L = left; R = right.

**Table 3.** Clusters of the between-group meta-analysis.

| **Map** | **Direction** | **Peak region** | **Peak MNI** | **Voxels (mm³)** | **SDM-*Z*** | ***g* or coefficient [95% CI]** | $p_{FWE}$ | **I² (%)** |
|---|---|---|---|---|---|---|---|---|
| Intercept | autistic > non-autistic | R TP | 42, 10, −20 | 422 (3,376) | 4.19 | 0.35 [0.19, 0.52] | 0.004 | 9.6 |
| Intercept | autistic > non-autistic | L SFG | −22, 6, 54 | 67 (536) | 5.14 | 0.44 [0.28, 0.61] | 0.015 | 25.1 |
| Intercept | autistic > non-autistic | R MTG | 60, −42, −10 | 46 (368) | 5.37 | 0.44 [0.28, 0.60] | 0.015 | 10.6 |
| Intercept | autistic > non-autistic | R PCUN | 16, −74, 44 | 21 (168) | 4.33 | 0.35 [0.19, 0.51] | 0.038 | 0.1 |
| Intercept | non-autistic > autistic | L mSTG | −58, −16, 12 | 5,490 (43,920) | −5.76 | −0.47 [−0.63, −0.31] | 0.001 | 2.7 |
| Intercept | non-autistic > autistic | L CAU | −14, 12, 6 | 232 (1,856) | −4.91 | −0.39 [−0.54, −0.23] | 0.003 | 2.6 |
| Intercept | non-autistic > autistic | R SFG | 30, 50, 20 | 30 (240) | −5.07 | −0.38 [−0.53, −0.24] | 0.027 | 6.2 |
| Intercept | non-autistic > autistic | R CAU | 10, 12, 8 | 25 (200) | −4.98 | −0.40 [−0.56, −0.24] | 0.027 | 5.5 |
| Intercept | non-autistic > autistic | L SMA | −4, 6, 60 | 13 (104) | −4.74 | −0.39 [−0.55, −0.23] | 0.039 | 3.0 |
| Modality | difference higher for written language | L mSTG | −50, −8, 2 | 293 (2,344) | −2.11 | −0.35 [−0.67, −0.02] | 0.003 | 1.1 |
| Modality | difference higher for written language | L PCC | −2, −40, 24 | 264 (2,112) | −2.34 | −0.39 [−0.72, −0.06] | 0.003 | 6.0 |
| Age | difference rises with age | R mSTG | 58, −8, 0 | 819 (6,552) | 2.54 | 0.028 [0.006, 0.050] | 0.002 | 15.6 |

| Map | Direction | Peak region | Peak MNI | Voxels (mm³) | SDM-$Z$ | $g$ or coefficient [95% CI] | $p_{FWE}$ | $I^2$ (%) |
|---|---|---|---|---|---|---|---|---|
| Age | difference rises with age | R CUN | 14, −78, 40 | 406 (3,248) | 2.61 | 0.027 [0.007, 0.048] | 0.002 | 0.3 |

*Note.* Positive values: autistic > non-autistic, or a moderator shifting the difference toward autistic > non-autistic. TFCE family-wise-error $p < 0.05$ (1,000 permutations), at least 10 voxels on a 2-mm grid. Values are at the cluster peak (largest |SDM-$Z$|). For the intercept, the effect is Hedges' $g$; for the moderators, it is the coefficient: the difference in $g$ between spoken and written language (modality) or the change in $g$ per year of mean age (age). $I^2$ is at the peak. a/m/pSTG and a/m/pSTS = anterior/mid/posterior superior temporal gyrus and sulcus; MTG = middle temporal gyrus; TP = temporal pole; SFG = superior frontal gyrus; SMA = supplementary motor area; PCC = posterior cingulate cortex; PCUN = precuneus; CUN = cuneus; CAU = caudate nucleus.

### 3.4. Modality

In both groups, brain activity was higher in spoken than in written studies in the bilateral superior temporal cortex (Figure 4, left and middle; Table 2; Supplementary Table S2). In the autistic group, these clusters lay in the right Rolandic operculum and STG (51,184 $mm^3$; coefficient 0.71, 95% CI 0.45 to 0.96) and the left posterior STG (22,592 $mm^3$; coefficient 0.60, 95% CI 0.33 to 0.87), and a small cluster in the left temporal pole was also higher in spoken studies (152 $mm^3$; coefficient 0.34, 95% CI 0.09 to 0.59). In the non-autistic group, they lay in the right mid STS (31,800 $mm^3$; coefficient 0.59, 95% CI 0.35 to 0.83) and the left MTG (31,440 $mm^3$; coefficient 0.83, 95% CI 0.47 to 1.19), and brain activity in the right SMA was higher in written studies (2,264 $mm^3$; coefficient −0.57, 95% CI −0.87 to −0.26).

Spoken studies showed a larger autistic reduction than written studies in the left mid STG (2,344 $mm^3$; coefficient −0.35, 95% CI −0.67 to −0.02), in a cluster that lay mostly within the left perisylvian reduction (Table 3; Figure 4A, right). The posterior cingulate cortex showed a group-by-modality interaction in the same direction (2,112 $mm^3$; coefficient −0.39, 95% CI −0.72 to −0.06). At the left mid STG peak, the predicted group difference was −0.52 (95% CI −0.73 to −0.32) for spoken and −0.07 (95% CI −0.32 to 0.18) for written studies; at the posterior cingulate peak, it was −0.16 (95% CI −0.38 to 0.07) for spoken and 0.24 (95% CI 0.00 to 0.49) for written studies (Figure 4B, right; Supplementary Table S2).

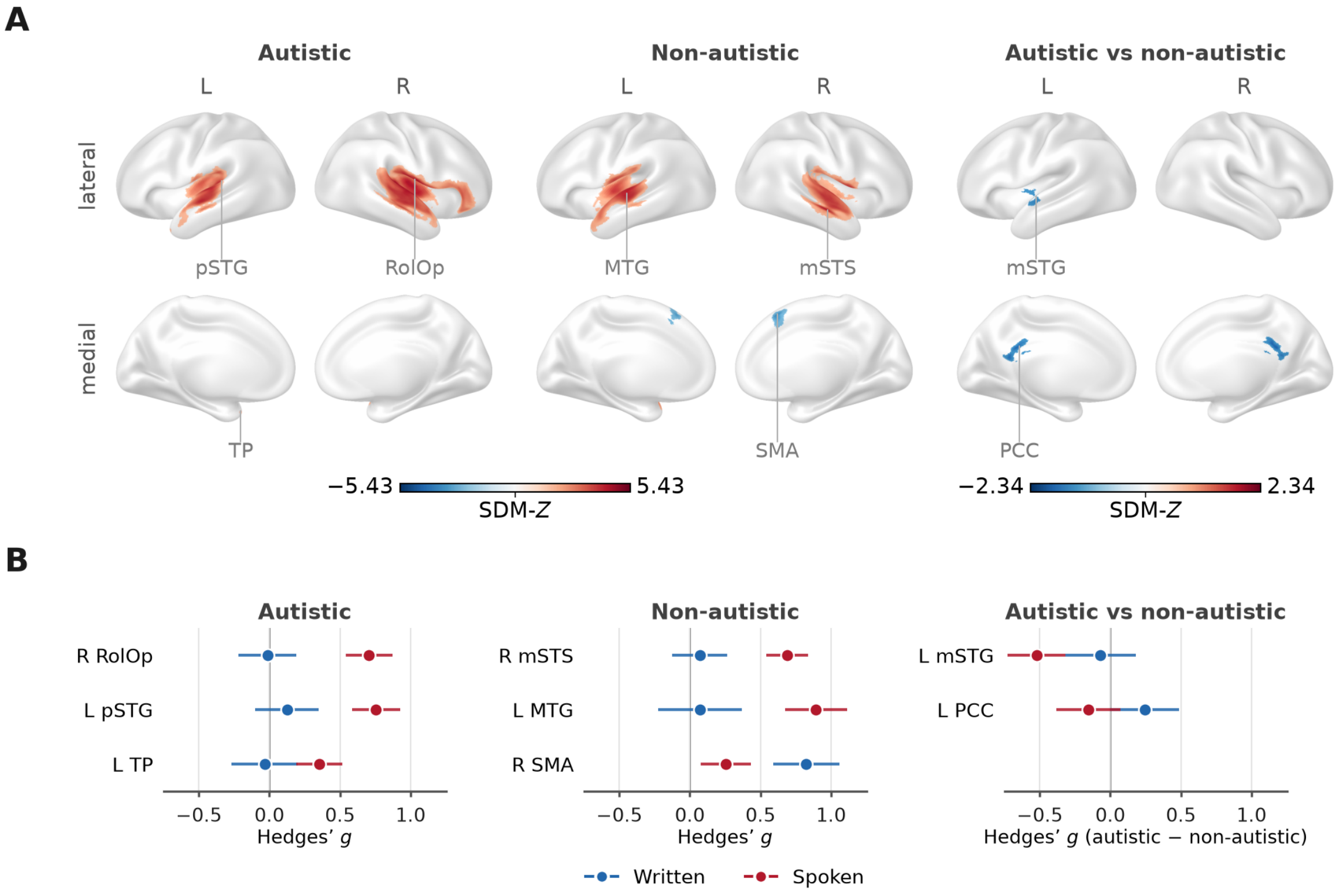

**Figure 4.** Effect of modality on brain activity in autistic and non-autistic groups and on the group difference. (A) The change in brain activity, or in the group difference, from written to spoken language. Red: higher for spoken language; blue: higher for written language. pSTG = posterior superior temporal gyrus; RolOp = Rolandic operculum; MTG = middle temporal gyrus; mSTS = mid superior temporal sulcus; TP = temporal pole; SMA = supplementary motor area; mSTG = mid superior temporal gyrus; PCC = posterior cingulate cortex; L = left; R = right. Axial slices in Supplementary Figure S2. (B) The effect the model predicts for written language (blue) and for spoken language (red) at the peak of every cluster of the maps in A, with its 95% confidence interval and age at its mean (Supplementary Table S2).

### 3.5. Age

Autistic brain activity rose with age in the left anterior insula (3,344 mm³; 0.044 *g* per year) and in the left cerebellum (lobule VI; 3,112 mm³; 0.035 per year), and fell with age in the right anterior STG (3,584 mm³; −0.024 per year) and in a small left temporal pole cluster (128 mm³; −0.024 per year). Non-autistic brain activity rose with age in the right cerebellum (lobule VIII; 29,512 mm³; 0.034 per year; Figure 5A, left and middle; Supplementary Figure S3; Table 2). At the three peaks that rose with age, brain activity was between 0.04 and 0.08 at the mean age of the pediatric samples of each meta-analysis and between 0.49 and 0.64 at that of the adult samples. At the two autistic clusters that fell with age, it was $g$ = 0.35 at the pediatric and $g$ =

0.04 at the adult mean age in the right anterior STG and $g = 0.37$ and $g = 0.08$ in the left temporal pole (Figure 5B, left and middle; Supplementary Table S2).

The right mid STG (6,552 $mm^3$; 0.028 *g* per year, 95% CI 0.006 to 0.050) and the right cuneus (3,248 $mm^3$; 0.027 per year, 95% CI 0.007 to 0.048) showed a group-by-age interaction, toward higher autistic than non-autistic activity in older samples (Table 3; Figure 5A, right). No cluster showed the opposite age effect. Over the age range of the samples, a slope of 0.028 per year amounts to a change of about 0.75 in *g*. The right cuneus cluster lay mostly outside both groups' activation maps (345 of its 406 voxels). At the right STG peak, the predicted group difference was −0.06 (95% CI −0.30 to 0.18) at the mean age of the pediatric samples (12.8 years) and 0.34 (95% CI 0.11 to 0.57) at that of the adult samples (26.9 years); at the right cuneus peak, it was 0.09 for pediatric samples (95% CI −0.11 to 0.29) and 0.52 for adult samples (95% CI 0.30 to 0.75; Figure 5B, right; Supplementary Table S2). In both regions, the model thus predicted higher autistic than non-autistic activity at the adult mean age (95% CIs above zero) and no detectable difference at the pediatric mean age (95% CIs including zero). Neither group's own meta-analysis showed an age slope at these two peaks: at the right mid STG peak, the autistic slope was −0.006 *g* per year (95% CI −0.032 to 0.020) and the non-autistic slope −0.003 (95% CI −0.034 to 0.028); at the right cuneus peak, they were 0.001 (95% CI −0.015 to 0.017) and 0.000 (95% CI −0.015 to 0.014).

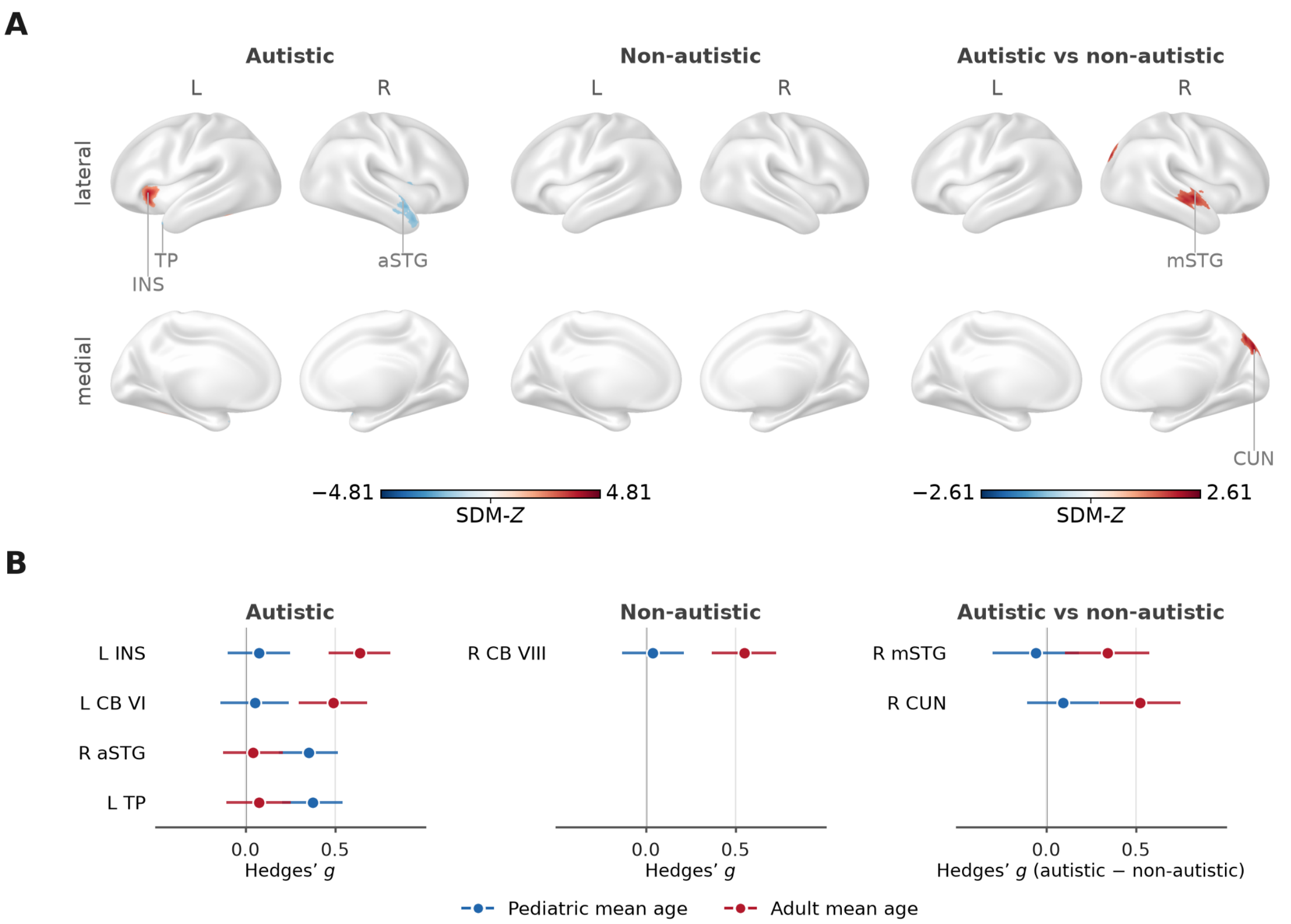

**Figure 5.** Effect of age on brain activity in autistic and non-autistic groups and on the group difference. (A) The change in brain activity, or in the group difference, per year of mean sample age. Red: rising with age; blue: falling with age. INS = insula; TP = temporal pole; aSTG = anterior superior temporal gyrus; mSTG = mid superior temporal gyrus; CUN = cuneus; CB VI and CB VIII = cerebellar lobules VI and VIII; L = left; R = right. Axial slices in Supplementary Figure S3. (B) The effect the model predicts at the mean age of the pediatric samples (blue) and of the adult samples (red) of each meta-analysis at the peak of every cluster of the maps in A, with its 95% confidence interval and modality at its mean (Supplementary Table S2).

## 4. Discussion

We synthesized 45 fMRI and PET studies of language processing in autism by SDM-PSI meta-regression, estimating each group's activation and the group difference and testing whether these effects depended on modality and age. Autistic and non-autistic brains activated a similar left-lateralized language network. Autistic activity was reduced throughout the left perisylvian cortex and increased in the right temporal pole. Spoken studies showed a larger autistic reduction than written studies in the left mid STG, and the posterior cingulate cortex showed a group-by-modality interaction in the same direction. The group difference in the right STG and the right cuneus widened with age, toward higher autistic activity in older individuals.

**4.1. Reduced autistic brain activity across the left perisylvian network**

In both groups, the language network peaked in the left IFG pars triangularis and spanned the perisylvian cortex of both hemispheres. The largest difference between the groups was a reduction of autistic activity within this network that spanned the left perisylvian cortex, from the IFG pars triangularis through the insula and operculum to the mid and posterior STG and the posterior STS. Both groups activate this cortex for language, so the reduction is a difference in degree within the typical language network. It agrees with the left inferior frontal reductions reported by Duvall et al. (2023) and Herringshaw et al. (2016) and with the left superior temporal reductions reported by these syntheses, by Cheng et al. (2026) and by Wu et al. (2026) in a region of interest. The reduction was larger in spoken than in written studies in the left mid STG, and no significant age effect on the group difference lay in this cortex. Reduced left perisylvian activity is thus one of the most consistent findings on language processing in autism. Most previous syntheses found it in the temporal cortex despite their different subdomains and age ranges, and here it was the largest group difference, with low heterogeneity at its peak ($I^2$ = 2.7%).

Based on the underconnectivity theory (Just et al., 2004, 2012), we expected reduced autistic activity in the left inferior frontal cortex and increased activity in the left posterior temporal cortex. The frontal prediction holds, but the posterior temporal cortex shows reduced autistic activity as well, so the data do not support a shift from frontal to posterior processing; they show reduced activity in both regions.

**4.2. Increased autistic brain activity in the right temporal cortex**

Reduced autistic brain activity lay mainly in the left hemisphere and increased autistic activity mainly in the right, which agrees with the reduced leftward lateralization of language in autism (Jouravlev et al., 2020; Lindell & Hudry, 2013). This pattern was clearest in the temporal cortex, where the left hemisphere showed reductions while the right hemisphere showed increases. The exceptions lay in the frontal cortex and the caudate nucleus: small reductions in the right superior frontal gyrus and the right caudate nucleus and a small increase in the left superior frontal gyrus. The main increase lay in the right temporal pole and the adjacent anterior insula. It agrees with Herringshaw et al. (2016), who found increased autistic activity in the right anterior STG (their peak at MNI 56, 12, −6) 3 mm from a local maximum of our temporal pole cluster (56, 14, −8).

The small left superior frontal increase (peak at −22, 6, 54) lies 8 mm from the left middle frontal increase they reported (−24, −2, 54). We found no increase in the right IFG, where they also reported one.

Jouravlev et al. (2020) found reduced lateralization to be driven by stronger right-hemisphere activity; we found it mainly as a left-hemisphere reduction, with a smaller right anterior temporal increase. Kana et al. (2006) found increased autistic activity in parietal and occipital regions associated with visual imagery. We found such an increase only in the right precuneus and cuneus: in the small precuneus cluster and, in adult samples, in the larger cuneus cluster that contains the precuneus cluster and whose group difference widened with age. The left superior frontal and right precuneus increases and the right superior frontal and right caudate reductions lay mostly outside both groups' language networks and may reflect group differences in deactivation rather than in activation.

### 4.3. A larger reduction for spoken language in the left superior temporal cortex

Based on the enhanced perceptual functioning account (Mottron et al., 2006), we expected a group difference that depends on the perceptual properties of the language. Compared with written studies, spoken studies showed a larger autistic reduction in the left mid STG within the perisylvian reduction, and the posterior cingulate cortex showed a group-by-modality interaction in the same direction. In the superior temporal cortex, brain activity was higher in spoken than in written studies in both groups, as expected from the auditory analysis of speech (Price, 2012). The larger autistic reduction for heard language thus lay in the auditory cortex. This is the direction the account predicts. Speech is a sound with complex temporal and spectral structure, the kind of stimulus for which the account proposes reduced autistic perception (Mottron et al., 2006). For temporally complex non-speech sounds, autistic participants showed reduced activity in the non-primary auditory cortex of the STG (Samson et al., 2011). Written words, in contrast, reach the language network as static visual forms through the visual cortex (Price, 2012). At its peak, the predicted group difference was $g = -0.52$ (95% CI −0.73 to −0.32) for spoken and $g = -0.07$ (95% CI −0.32 to 0.18) for written studies; the reduction was thus detectable for spoken studies only, where its 95% CI excluded zero.

### 4.4. A group difference in the right temporal and occipital cortex that widened with age

The right STG and the right cuneus showed a group-by-age interaction: the model predicted higher autistic than non-autistic activity there at the mean age of the adult samples (right STG: $g = 0.34$; right cuneus: $g = 0.52$) and no detectable difference at that of the pediatric samples. Taken at face value, this is the direction expected if a group difference in lateralization grows with age (Flagg et al., 2005). Two earlier syntheses found reduced autistic superior temporal activity in pediatric but not in adult studies: in the right STG across social tasks (Dickstein et al., 2013) and in the bilateral STG across auditory and language tasks (Philip et al., 2012). The right STG interaction has the same direction, but the left perisylvian reduction showed no significant age effect. We are cautious, however, in reading it as developmental change. Neither group's own meta-analysis showed an age slope at either peak, so the effect is not traceable to a change in either group's activation. Future studies that compare pediatric and adult groups within one design, or follow the same participants from childhood into adulthood, could test whether the higher autistic activity in the right STG and cuneus emerges with age and whether the left perisylvian reduction is stable from childhood into adulthood.

### 4.5. Conclusion

Autistic language processing relies on the typical left-lateralized language network, with reduced activity within it rather than a shift to other regions. This reduction was the most extensive group difference and was larger for spoken than for written language. Increases in autistic activity were less extensive and lay mainly in the right hemisphere. These group differences thus show reduced leftward lateralization of language in autism.

## Acknowledgements

We thank Nan Wang for her insights and helpful comments.

**Studies included in the meta-analyses**

The 45 studies included in the meta-analyses are listed here; each study is cited in Table 1.

## Supplementary Materials

# Language processing in the autistic brain:
# A meta-analysis of neuroimaging studies across modalities and age

## Supplementary Material S1. Search, selection and extraction

### S1.1 Databases and search strings

Searches were run on 4 January 2026 for records published between 1 January 1996 and 31 December 2025, with review articles excluded by document type. The strings below are the database queries.

**Web of Science (1,091 records)**

```
TS=(autism OR autistic OR ASD OR ASC OR Asperger*) AND TS=(language OR
speech OR verbal OR linguistic* OR text OR reading OR listening OR word*
OR lexic* OR phrase* OR sentence* OR discourse OR narrative* OR
communicat* OR semantic* OR syntax OR syntactic* OR morpholog* OR
morphosynta* OR phonolog* OR phonetic* OR prosod* OR orthograph* OR
pragmatic*) AND TS=(comprehension OR processing OR understanding OR
perception OR recognition OR identification OR encod* OR decod*) AND
TS=(neuroimaging OR "functional magnetic resonance imaging" OR "functional
MRI" OR fMRI OR "positron emission tomography" OR PET OR "functional
connectivity" OR BOLD OR hemodynamic) NOT DT=(Review)
```

**PubMed (1,352 records)**

```
(autism OR autistic OR ASD OR ASC OR Asperger*) AND (language OR speech OR
verbal OR linguistic* OR text OR reading OR listening OR word* OR lexic*
OR phrase* OR sentence* OR discourse OR narrative* OR communicat* OR
semantic* OR syntax OR syntactic* OR morpholog* OR morphosynta* OR
phonolog* OR phonetic* OR prosod* OR orthograph* OR pragmatic*) AND
(comprehension OR processing OR understanding OR perception OR recognition
OR identification OR encod* OR decod*) AND (neuroimaging OR "functional
magnetic resonance imaging" OR "functional MRI" OR fMRI OR "positron
emission tomography" OR PET OR "functional connectivity" OR BOLD OR
hemodynamic) NOT ("Review"[Publication Type])
```

**Scopus (1,284 records exported; 1,298 returned on the website)**

```
TITLE-ABS-KEY(autism OR autistic OR ASD OR ASC OR Asperger*) AND TITLE-
ABS-KEY(LANGUAGE OR speech OR verbal OR linguistic* OR text OR reading OR
listening OR word* OR lexic* OR phrase* OR sentence* OR discourse OR
narrative* OR communicat* OR semantic* OR syntax OR syntactic* OR
morpholog* OR morphosynta* OR phonolog* OR phonetic* OR prosod* OR
orthograph* OR pragmatic*) AND TITLE-ABS-KEY(comprehension OR processing
OR understanding OR perception OR recognition OR identification OR encod*
OR decod*) AND TITLE-ABS-KEY(neuroimaging OR "functional magnetic
resonance imaging" OR "functional MRI" OR fMRI OR "positron emission
tomography" OR PET OR "functional connectivity" OR BOLD OR hemodynamic)
AND NOT DOCTYPE(re)
```

## S1.2 Record flow

The record flow is shown in Figure 1. Duplicates were removed by DOI or PubMed ID match (1,117) and by manual check (227). After title and abstract screening, 55 reports from the database search were assessed in full text, together with eight reports that only the check of previous syntheses found. Of the 63 reports, 45 were included; Supplementary Table S1 lists the 18 excluded reports with their primary reason. The previous syntheses screened were Cheng et al. (2026), Herringshaw et al. (2016), Hua et al. (2024), Larson et al. (2025), Li et al. (2022), Phan et al. (2021) and Tryfon et al. (2018); Supplementary Table S1b lists the syntheses that include or cite each of these eight reports.

**Supplementary Table S1.** Reports excluded at full-text screening, with the primary reason (the first that applied, in the order of Figure 1) and the source of the record.

| Report | Source | Primary reason |
|---|---|---|
| Just et al., 2004 | Database | No mean age reported |
| Redcay & Courchesne, 2008 | Database | Participants scanned asleep |
| Knaus et al., 2010 | Previous syntheses | Results only in regions of interest or masks |
| Colich et al., 2012 | Database | Results only in regions of interest or masks |
| Eyler et al., 2012 | Database | Participants scanned asleep |
| Watanabe et al., 2012 | Database | No eligible contrast |
| Kenworthy et al., 2013 | Database | Results only in regions of interest or masks |
| Lo et al., 2013 | Previous syntheses | Results only in regions of interest or masks |
| Moseley et al., 2013 | Database | No coordinates |

| Report | Source | Primary reason |
|---|---|---|
| Radulescu et al., 2013 | Database | No coordinates |
| Moseley et al., 2014 | Previous syntheses | Not fMRI or PET |
| Verly et al., 2014 | Database | No coordinates |
| Jochaut et al., 2015 | Database | Results only in regions of interest or masks |
| Sharda et al., 2015 | Database | Results only in regions of interest or masks |
| Eigsti et al., 2016 | Database | Activation only pooled across groups |
| Bednarz et al., 2017 | Previous syntheses | Results only in regions of interest or masks |
| Wong et al., 2019 | Database | No eligible contrast |
| Xu et al., 2025 | Database | Group difference in one direction only |

**Supplementary Table S1b.** The eight reports found only through previous syntheses, with the syntheses that include or cite each.

| Report | Status | Syntheses that include or cite it |
|---|---|---|
| Bednarz et al., 2017 | Excluded | Larson et al., 2025; Phan et al., 2021 |
| Doyle-Thomas et al., 2013 | Included | Hua et al., 2024; Tryfon et al., 2018 |
| Gebauer et al., 2014 | Included | Tryfon et al., 2018 |
| Green et al., 2018 | Included | Hua et al., 2024 |
| Knaus et al., 2010 | Excluded | Li et al., 2022; Phan et al., 2021 |
| Lo et al., 2013 | Excluded | Phan et al., 2021 |
| Moseley et al., 2014 | Excluded | Herringshaw et al., 2016 |
| Schelinski et al., 2016 | Included | Tryfon et al., 2018 |

### S1.3 Statistics and thresholds

Each study's threshold for SDM-PSI was taken from its voxel-level or cluster-forming threshold. SPM and FSL $p$ values were read as one-tailed and AFNI and BrainVoyager $p$ values as two-tailed. Where a study's threshold was FWE- or FDR-corrected or not stated, $p = 0.001$ was used. The threshold was never set above the study's smallest printed peak $t$.

**Supplementary Figures S1–S3. The maps of Figures 2, 4 and 5 as axial slices**

Thresholds and color scale: TFCE family-wise-error $p < 0.05$ (1,000 permutations), at least 10 voxels, signed SDM-*Z* on a dark blue to dark red scale that is symmetric about zero (white), with one range for panels A and B of each figure, red for positive and blue for negative values. Slices are on the MNI152 template, with the left hemisphere on the left. The between-group intercept map is shown on slices in Figure 2B. Labels as in Tables 2 and 3.

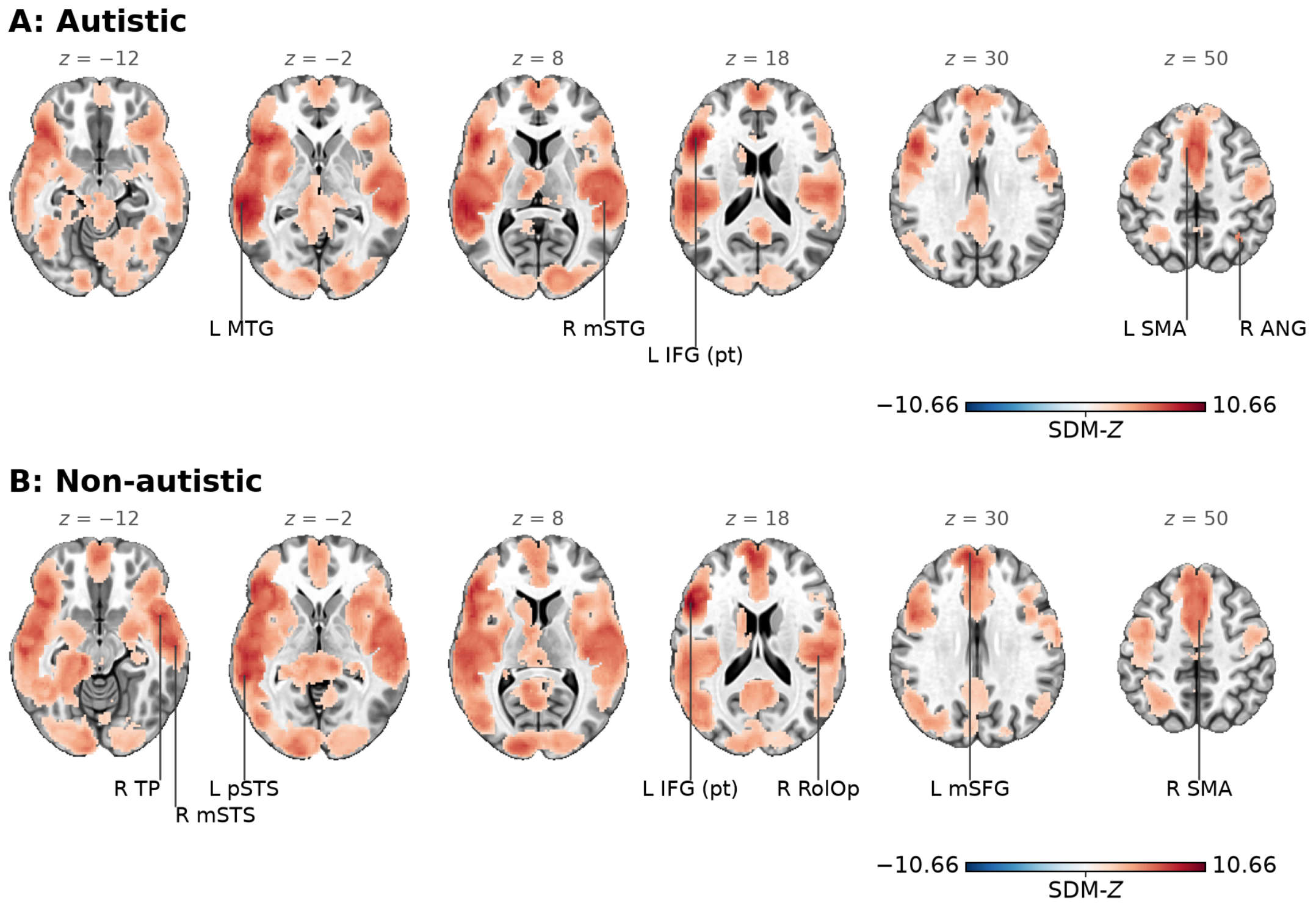

**Supplementary Figure S1.** Within-group intercept maps (Figure 2A, left and middle). (A) Autistic, language > control; (B) non-autistic, language > control.

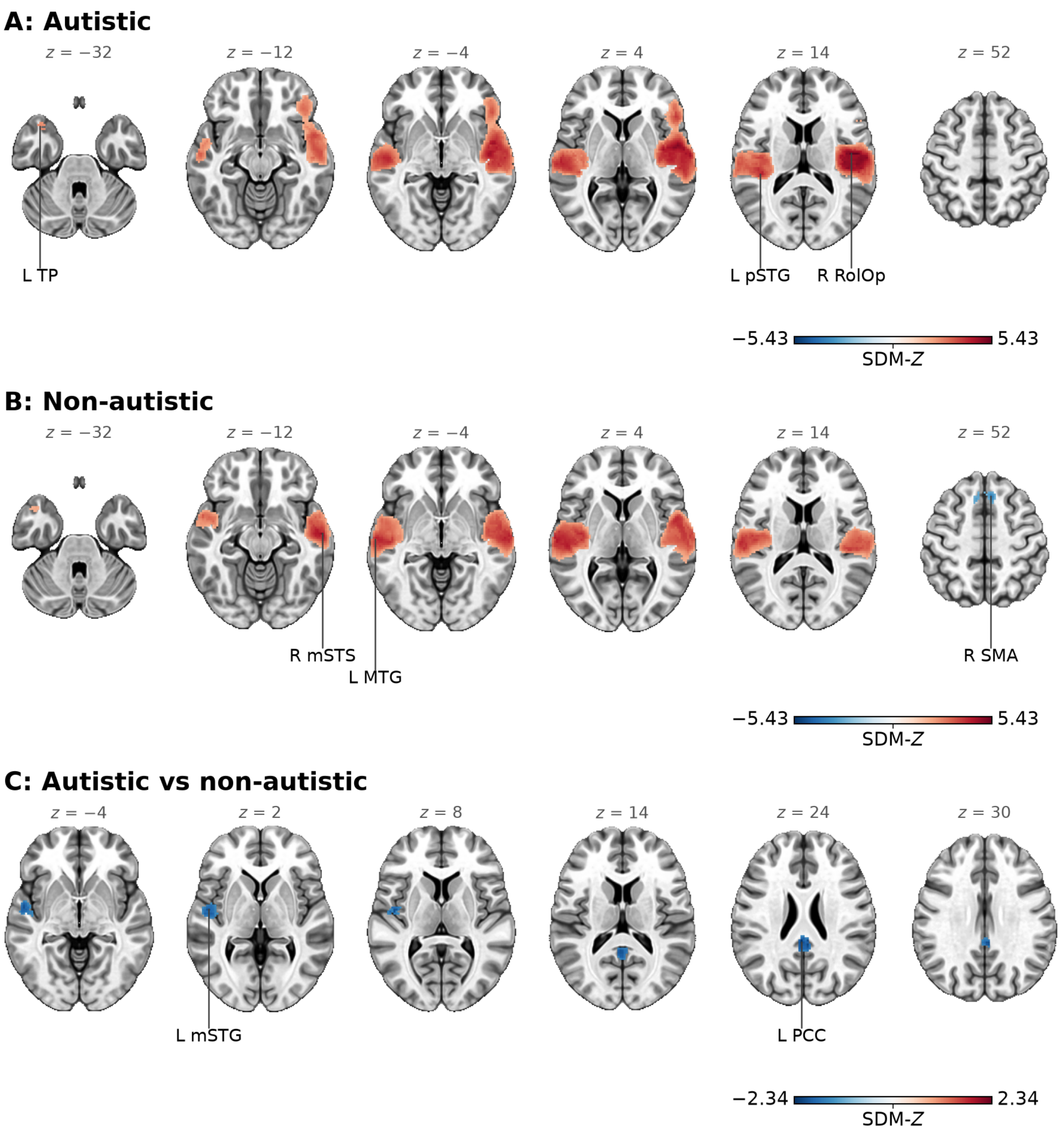

**Supplementary Figure S2.** Modality (Figure 4). (A) Autistic; (B) non-autistic; (C) autistic vs non-autistic.

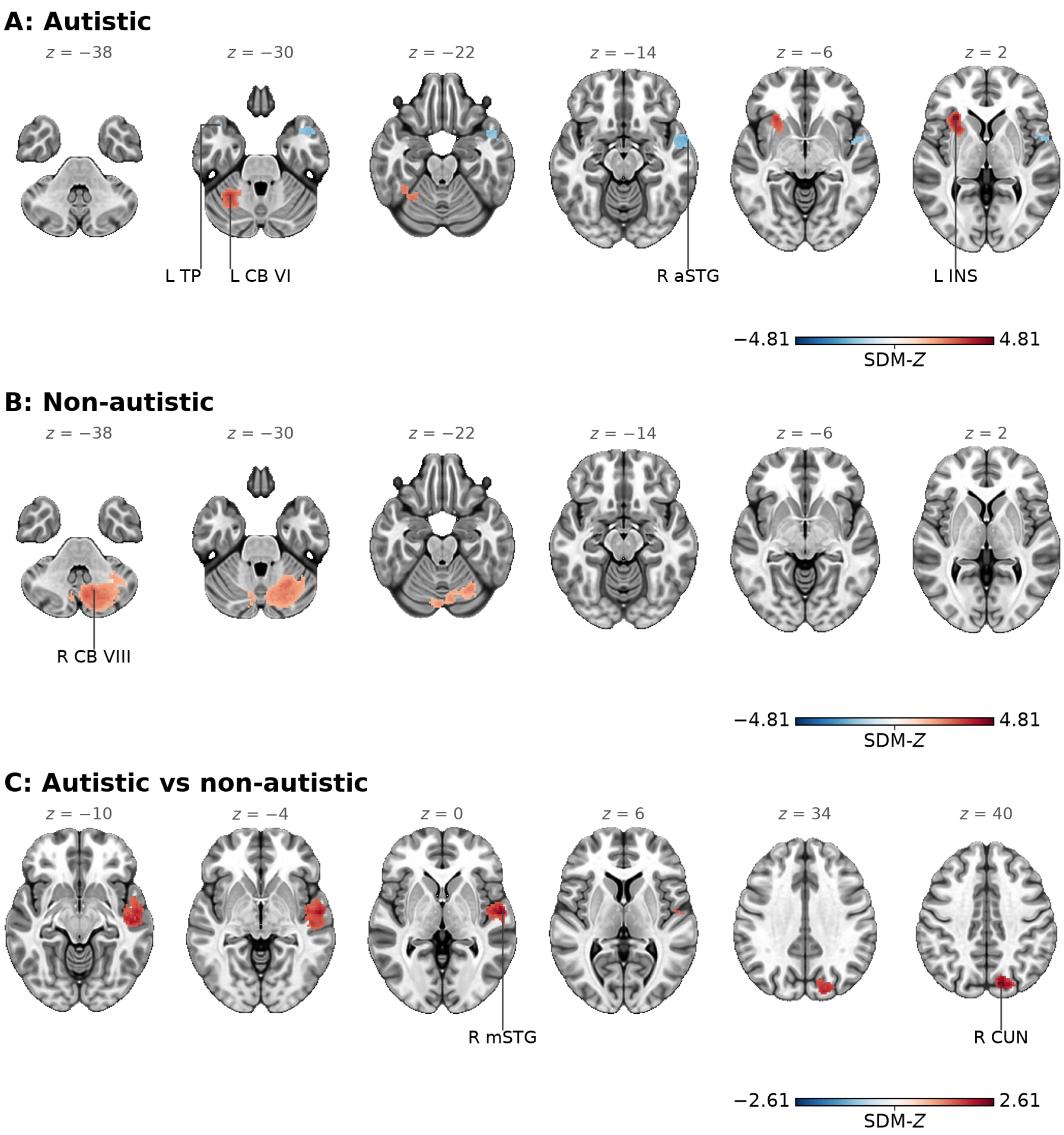

**Supplementary Figure S3.** Age (Figure 5). (A) Autistic; (B) non-autistic; (C) autistic vs non-autistic.

## Supplementary Figure S4. Funnel plots

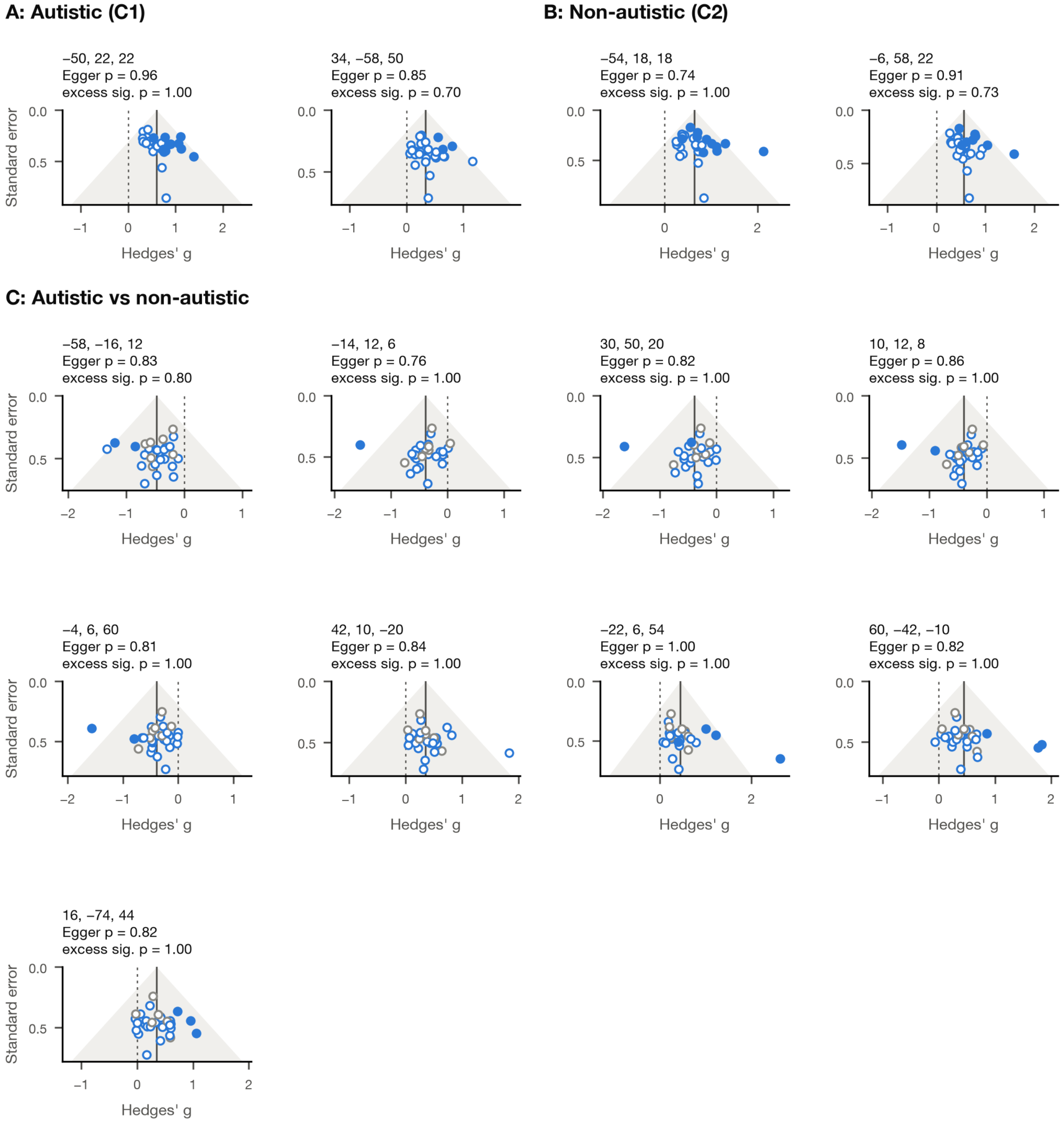

**Supplementary Figure S4.** Funnel plots at the peaks of the intercept maps: (A) autistic, (B) non-autistic, (C) autistic vs non-autistic. Each point is one study's Hedges' $g$ at the peak, combined over the 50 imputations by Rubin's rules, against its standard error. Filled: a study with a reported peak within 10 mm of the peak; open blue: a study whose nearest reported peak lies more than 10 mm from the peak; open gray: a null study. Shaded: pseudo 95% region round the pooled $g$ (solid line); dotted line: $g = 0$. The Egger-type test of SDM-PSI uses the model residuals.

## Supplementary Figure S5 and Table S2. Marginal means

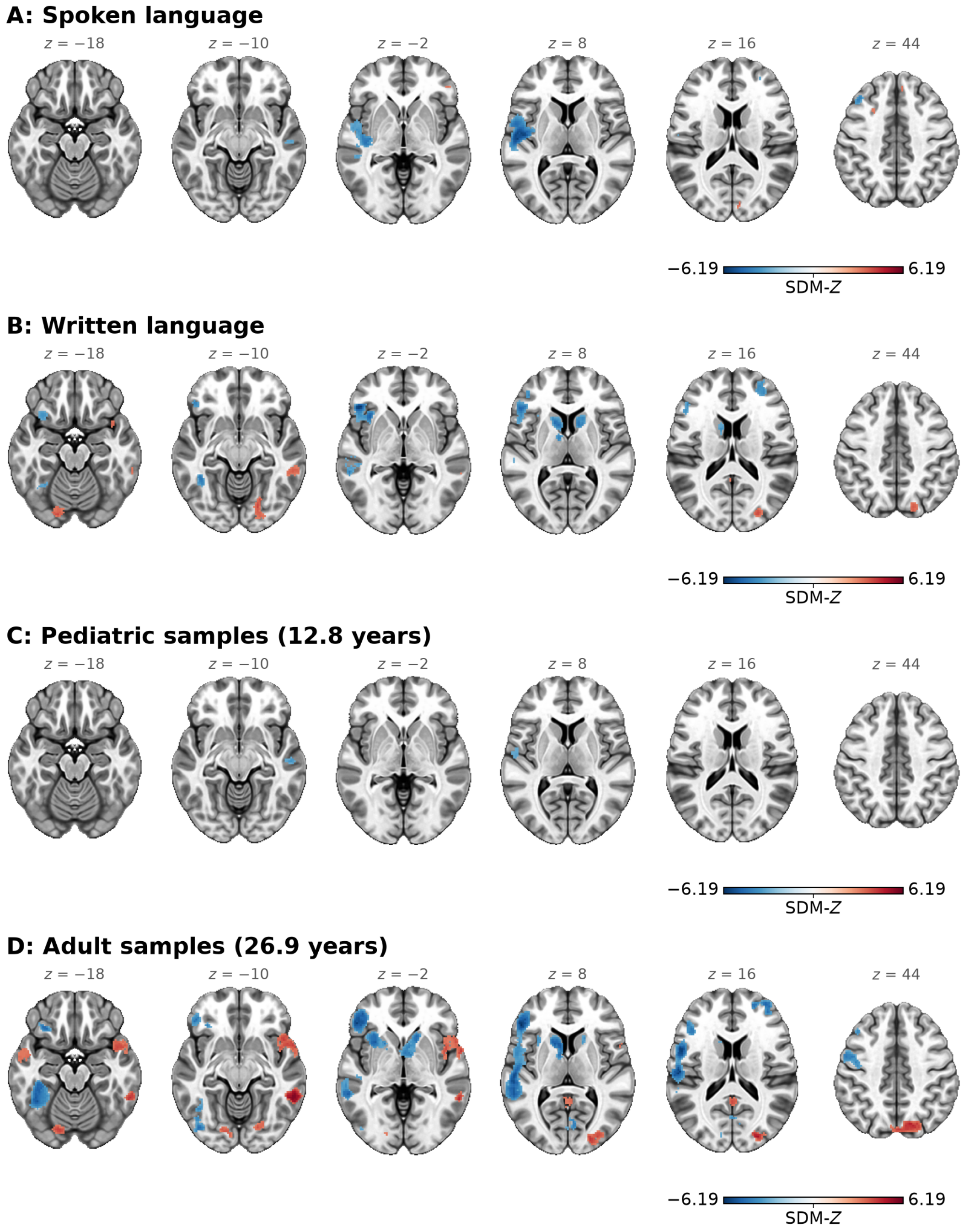

**Supplementary Figure S5.** Between-group level maps: the group difference (autistic − non-autistic) that the model predicts (A) for spoken and (B) for written language with age at its mean, and at the mean age of (C) the pediatric samples (12.8 years) and (D) the adult samples (26.9 years) with modality at its mean. Colors show SDM-*Z* (red: autistic > non-autistic; blue: non-autistic > autistic). These maps are thresholded at an uncorrected voxel-level $p < 0.001$ with at least 10 voxels and are descriptive.

**Supplementary Table S2.** Marginal means at the peak of every moderator cluster: the effect (Hedges' *g*, with 95% CI) that the model predicts at each level of the moderator, with age at its mean for the two modalities and modality at its mean for the two ages.

| Meta-analysis | Moderator | Peak region | Peak MNI | Coefficient | Level | *g* [95% CI] | Level | *g* [95% CI] |
|---|---|---|---|---|---|---|---|---|
| Autistic | Modality | R RolOp | 46, −12, 14 | 0.71 | Spoken | 0.70 [0.54, 0.87] | Written | −0.01 [−0.22, 0.19] |
| Autistic | Modality | L pSTG | −42, −32, 12 | 0.60 | Spoken | 0.75 [0.58, 0.93] | Written | 0.12 [−0.10, 0.35] |
| Autistic | Modality | L TP | −38, 14, −32 | 0.34 | Spoken | 0.35 [0.19, 0.52] | Written | −0.03 [−0.27, 0.20] |
| Autistic | Age | L INS | −30, 24, 2 | 0.044 | Pediatric (13.5 years) | 0.08 [−0.10, 0.25] | Adult (26.1 years) | 0.64 [0.47, 0.81] |
| Autistic | Age | L CB VI | −30, −52, −30 | 0.035 | Pediatric (13.5 years) | 0.05 [−0.14, 0.24] | Adult (26.1 years) | 0.49 [0.30, 0.68] |
| Autistic | Age | R aSTG | 62, 0, −14 | −0.024 | Pediatric (13.5 years) | 0.35 [0.19, 0.52] | Adult (26.1 years) | 0.04 [−0.12, 0.21] |
| Autistic | Age | L TP | −38, 16, −32 | −0.024 | Pediatric (13.5 years) | 0.37 [0.21, 0.54] | Adult (26.1 years) | 0.08 [−0.10, 0.25] |
| Non-autistic | Modality | R mSTS | 60, −12, −12 | 0.59 | Spoken | 0.69 [0.54, 0.84] | Written | 0.07 [−0.12, 0.26] |
| Non-autistic | Modality | L MTG | −64, −18, 0 | 0.83 | Spoken | 0.89 [0.67, 1.11] | Written | 0.07 [−0.22, 0.37] |
| Non-autistic | Modality | R SMA | 6, 24, 52 | −0.57 | Spoken | 0.25 [0.08, 0.43] | Written | 0.83 [0.59, 1.06] |
| Non-autistic | Age | R CB VIII | 12, −68, −48 | 0.034 | Pediatric (13.0 years) | 0.04 [−0.13, 0.21] | Adult (26.8 years) | 0.55 [0.37, 0.73] |

| **Meta-analysis** | **Moderator** | **Peak region** | **Peak MNI** | **Coefficient** | **Level** | ***g* [95% CI]** | **Level** | ***g* [95% CI]** |
|---|---|---|---|---|---|---|---|---|
| Autistic vs non-autistic | Modality | L mSTG | −50, −8, 2 | −0.35 | Spoken | −0.52 [−0.73, −0.32] | Written | −0.07 [−0.32, 0.18] |
| Autistic vs non-autistic | Modality | L PCC | −2, −40, 24 | −0.39 | Spoken | −0.16 [−0.38, 0.07] | Written | 0.24 [0.00, 0.49] |
| Autistic vs non-autistic | Age | R mSTG | 58, −8, 0 | 0.028 | Pediatric (12.8 years) | −0.06 [−0.30, 0.18] | Adult (26.9 years) | 0.34 [0.11, 0.57] |
| Autistic vs non-autistic | Age | R CUN | 14, −78, 40 | 0.027 | Pediatric (12.8 years) | 0.09 [−0.11, 0.29] | Adult (26.9 years) | 0.52 [0.30, 0.75] |

*Note.* Age levels are the mean ages of the pediatric (under 18 years) and the adult samples of each meta-analysis. In the between-group rows, *g* is the autistic − non-autistic difference. The coefficient is that of the moderator map: per year for age, otherwise the difference between the levels.

## References

Works cited in this supplement are listed in the reference lists of the main article, except the following reports excluded at full text (Supplementary Tables S1 and S1b).